\documentclass[prc,aps,superscriptaddress,nofootinbib]{revtex4}
\usepackage{amscd}
\usepackage{amsmath}
\usepackage{bm}
\usepackage[dvips]{color}                 
\usepackage{graphicx}
\usepackage{eucal}
\usepackage{color}
\usepackage{colordvi}
\usepackage{axodraw}
\usepackage{subfigure}

\newcommand{\x}{{\mathbf{x}}}
\newcommand{\y}{{\mathbf{y}}}
\newcommand{\kv}{{\mathbf{k}}}
\newcommand{\qv}{{\mathbf{q}}}
\newcommand{\pv}{{\mathbf{p}}}
\newcommand{\Pv}{{\mathbf{P}}}

\newcommand{\beq}{\begin{equation}}
\newcommand{\eeq}{\end{equation}}
\newcommand{\bea}{\begin{eqnarray}}
\newcommand{\eea}{\end{eqnarray}}

\newcommand{\nn}{\nonumber}
\newcommand{\benn}{\begin{displaymath}}
\newcommand{\eenn}{\end{displaymath}}
\newcommand{\ket}[1]{| #1 \rangle}                     
\newcommand{\bra}[1]{\langle #1 \, |}                  

\begin{document}
\leftline{January 28 2004}
\preprint{\vbox{
\hbox{LBNL-54557}
}}

\title{\bf \LARGE Aharonov-Bohm effect and nucleon-nucleon phase shifts on the lattice}

\author{Paulo F. Bedaque\footnote{{\tt pfbedaque@lbl.gov}}}
\affiliation{Lawrence-Berkeley Laboratory, Berkeley, CA 94720}

\begin{abstract}
We propose a method for the lattice QCD computation of nucleon-nucleon low-energy interactions. It consists in simulating QCD in the background of a ''electromagnetic"
field whose potential is non-vanishing, but whose field strength is zero. By tuning the background field, phase-shifts at any (but small) momenta can be determined by measuring the shift of the ground state energy. Lattice sizes as small as $5$ Fermi can be sufficient for the calculation of phase shifts up to momenta of order of $m_{\pi}/2$.
\end{abstract}
\maketitle
\bigskip

\maketitle

One of the central goals of nuclear physics is to relate the successful phenomenological models developed throughout the years with the underlying fundamental theory of the strong interactions, QCD. Effective field theories are an important step in this direction, but they are inherently limited by the existence of low energy constants whose values are not determined by symmetries and have to be fit to experiment. The need is then obvious for a fully non-perturbative method that can determine the interaction between nucleons (or alternatively, the low energy constants of the effective theory) directly from QCD. At present, lattice QCD is the only such method.   

Most phenomenological models of nuclei are based on non-relativistic two(and three) nucleon potentials. However, since nucleons are not infinitely heavy, the inter-nucleon potential is not a well defined quantity that can be measured on the lattice, even in principle. Instead, the connection between QCD and nuclear physics should be established through observables like scattering amplitudes and phase shifts, etc.. That brings out a problem: lattice calculations are done in euclidean space and analytic continuation of the euclidean correlation functions at infinite volume to Minkowski space is, in practice, impossible. This observation, formalized in \cite{maiani_testa}, seems to restrict lattice QCD to observables like masses, decays constants and amplitudes at kinematical thresholds. Phase shifts at some special values of the momenta can however be obtained by measuring the shifts in the low lying two-particle states due to the finite volume \cite{hamber_et_al,luscher1,luscher2}, as long as the lattice size $L$ is larger than the pion Compton wavelength (up to corrections of order $e^{-m_\pi L}$). 
This can be intuitively understood by realizing that the baryon number two sector of QCD at momenta smaller than the pion mass reduces to a non relativistic quantum mechanical system with two nucleons interacting through contact interactions. At momenta $Q$ much smaller than the $\sim 1/a$, where $a$ is the nucleon-nucleon scattering length, this contact interaction is perturbative but it becomes strong at $Q\sim 1/a$.
In particular, for lattices with size $L$ much larger than the scattering length  $a$ the low lying states have typical momenta $Q$ satisfying $Q<< 1/a$, and Luscher derived the formula relating the shifts in the energy levels and $a$ as an expansion in powers of $a/L$. This method has been used to obtain pion-pion scattering phase shifts \cite{pion_pion} but in the two nucleon sector I am aware of only only one quenched calculation performed with a large pion mass \cite{fukugita_et_al}.

In the two-nucleon case the condition $L \gg a$ can hardly be satisfied since the
scattering lengths between two nucleons are large by QCD standards ($5.42$ fm in the spin triplet and $23.7$ fm in the spin singlet channel) and numerical simulations with lattice sizes much larger than this are impractical. For $L \cong a$ the shifts in the energy levels due to the nucleon-nucleon interactions are not small but can still be reliably computed and used to obtain information on the nucleon-nucleon interactions.
An analysis of this method in the two-nucleon case was presented in \cite{nplqcd1}. There we found that, after taking into account the strong nucleon-nucleon interactions,  lattice sizes $L\sim 8$ fm are necessary for  the ground state to have small enough energy for the method to be valid, and even larger sizes if  the excited states are considered. From the shift in the ground state energy the phase shifts at only  one kinematical point can be determined. More handles on the phase shifts coming from the excited states would require even larger lattice sizes. 

This paper proposes a method that i) allows for smaller lattice sizes and ii) provides information about the phase shifts at any momenta smaller than $m_\pi/2$. The basic idea is very simple: one just simulates the baryon number two sector of QCD in a finite torus and in the background of a fictitious ''magnetic" potential with zero field strength, the kind of field generated by a thin solenoid going around inside the torus. 
Due to the Aharonov-Bohm effect \cite{ehrenberg_siday,aharonov_bohm}the energy levels are changed by this potential despite the fact that the field strength vanishes everywhere on the lattice. The strength of the potential can then be adjusted in order to have the ground state to have any energy desired. Alternatively we can describe the method as simulating QCD with twisted boundary conditions for the quark in one chosen spatial direction. The two descriptions are related by a change of variables amounting to a discontinuous gauge transformation.

\begin{figure}[t]
\includegraphics*[bb = 0 0 299 225, scale=0.7,clip=false]
{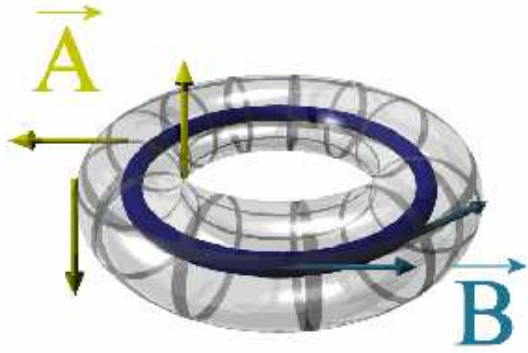}
\caption{\textit{The lattice with periodic boundary conditions (and two dimensions suppressed) is represented by the surface of the outer torus. The fictitious solenoid (inner ring) generates a magnetic vector potential $\vec{A}$ along direction $z$ (wrapped around the torus). The magnetic field is confined inside the ring and vanishes at the surface of the torus, where the lattice is.  } } 
\end{figure}

\subsection{QCD with the background field}

We will consider QCD in the presence of a $U(1)$ background gauge field coupling to baryon number of the form

\beq
\vec{A}=\frac{\phi}{3L}\hat{z},
\eeq where $\hat{z}$ is the unit vector in the $z$ direction and $\phi$ is real.

 In the case of two degenerate flavors of Wilson fermions the quark action  is

\begin{eqnarray}\label{qcd_action}
\mathcal {S}_q &=& \frac{1}{2b}\mathop{\sum}_{\x,\hat\mu}\bar q_\x\left[ (\gamma_\mu-r)\Omega_\mu(\x)q_{\x+\hat\mu}-(\gamma_\mu+r)\Omega_\mu^\dagger(\x-\hat\mu)q_{\x-\hat\mu}\right]+\frac{m_q b+4r}{b}\mathop{\sum}_{\x,\hat\mu}\bar q_\x q_\x\\
&\equiv& \mathop{\sum}_{\x,\y} \bar q_\x M_{\x\y}q_{\y},
\end{eqnarray} where $b$ is the lattice spacing,$i$ indexes the lattice sites, $\hat\mu$ the directions of the links,   $r$ is the Wilson term coefficient, $M_{ij}$ defines the quark operator and a sum over flavors is implicit. The link operators $\Omega_\mu(\x)$ are a product of a $SU(3)$ matrix $U_\mu(\x)$ and a phase determined by the background field $\Omega_\mu(\x)=U_\mu(\x) e^{ib A_\mu(\x)}$. 
Even in the presence o the background field the determinant of $M$ is positive. For that notice that the matrix $M$ satisfies $\gamma_5 M^\dagger_{\x\y}\gamma_5 = M_{\y\x}$ ,
 where the dagger means hermitian conjugation on the spin and colors indices only. This relation implies that 
$det(M)^*=det(M^\dagger)=det(\gamma_5 M \gamma_5)=det(M)$. The quark determinant, being a product of a up quark determinant and a down quark determinant, is then positive even in the presence of the background field and standard Monte Carlo techniques are available.

Instead of using the quark fields above, satisfying periodic boundary conditions, we can use instead

\begin{eqnarray}
\tilde q_\x &=& e^{i\frac{\x_z b\phi}{3L}} q_\x\nonumber\\
\bar{\tilde{q}}_\x &=& e^{-i\frac{\x_z b\phi}{3L}} \bar q_\x.
\end{eqnarray} The $\tilde q, \bar{\tilde{q}}$ fields satisfy twisted boundary conditions at $z=L$:
\begin{eqnarray}
\tilde{q}_{z=N} &=& e^{i\frac{\phi}{3}}\tilde{q}_{z=0}\nonumber\\
\bar{\tilde{q}}_{z=N} &=& e^{-i\frac{\phi}{3}}\tilde{\bar{q}}_{z=0},
\end{eqnarray} where $N$ is the number of sites in the $z$ direction.

We will extract the nucleon-nucleon phase shifts from the long (euclidean) time behavior of the finite volume correlator

\begin{eqnarray}\label{correlator}
C(t,\pv) &=&\bra{0}T N^T(\pv,t) \Pv N(-\pv,t) \ .\  N^\dagger(-\pv,0) \Pv^{\dagger} N^*(\pv,0) \ket{0}\nn\\
&&\mathop{\longrightarrow }_{t\rightarrow\infty}e^{-E_0 t} |\bra{E_0}  N^\dagger(-\pv,0) \Pv^{\dagger} N^*(\pv,0) \ket{0}|^2 ,
\end{eqnarray} where $P$ is the projector on the spin triplet or spin singlet  channels, $N(\kv,t)$ are operators with the quantum numbers of nucleons with momentum $\kv$ at time $t$ and $E_0$ is the energy of the ground state with the quantum numbers of the $N^\dagger(-\pv,t) \Pv^{\dagger} N^*(\kv,t) \ket{0}$ state. 

\subsection{The effective theory}

For small momenta $Q < m_\pi$, the nucleon-nucleon interaction can be described by an effective field theory containing only nucleons as explicit degrees of freedom. This effective theory has been used extensively in the computation of few-nucleon observables and has been reviewed in \cite{bedaque_bira_review, beane_et_al_review, phillips_review}. It contains only contact interactions, with increasing number of derivatives, and we will denote the coefficient of terms with $2n$ derivatives by $C_{2n}$. 
In the presence of the background field all  derivatives in the effective lagrangian are substituted by covariant derivatives. That is the {\it only} way that the background field can enter in the effective theory. Terms that are gauge invariant by themselves, for instance, anomalous magnetic terms, vanish since the magnetic field vanishes. We can perform a similar change of variables as above and eliminate the background field by working with fields satisfying twisted boundary conditions
\begin{eqnarray}\label{bc}
\psi(x,y,L) &=& e^{i\phi}\psi(x,y,0)\nonumber\\
\psi^\dagger(x,y,L) &=& e^{-i\phi}\psi^\dagger(x,y,0)
\end{eqnarray}

\begin{figure}[t]\label{bubbles}
\includegraphics*[bb = 83 413 507 525, scale=0.6,clip=false]
{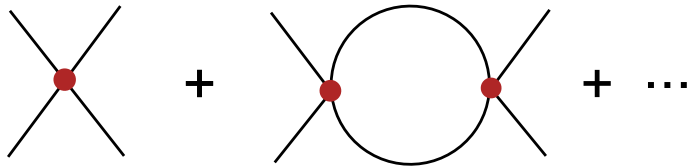}
\caption{\textit{Sum of graphs determining the tw0-nucleon scattering amplitude in the effective theory. The vertices include interactions with an arbitrary number of derivatives.  } } 
\end{figure}

For momenta $Q \ll 1/a$ the effective theory is perturbative but for $Q\sim 1/a$ it is non-perturbative. In fact, the two-nucleon scattering amplitude is given by the infinite series of diagrams shown in Fig.~\ref{bubbles}. Let us first consider the case of the spin singlet channel.  Using dimensional regularization the diagrams in Fig.~\ref{bubbles} can be computed and summed. The result is \cite{bedaque_bira_review, beane_et_al_review, phillips_review} 

\begin{eqnarray}\label{amplitude}
{\mathcal A} &=&\frac{\sum_n C_{2n}(\mu) k^{2n}}{1-I_0\sum_n C_{2n}(\mu) k^{2n} }\nonumber\\
&=&\frac{4\pi}{M}\frac{1}{k\cot \delta -ik} 
\end{eqnarray} where $k$ is the center-of-mass momentum of the colliding particles, $\mu$ is the renormalization point, $\delta$ the phase shift at momenta $k$and  the loop sum $I_0$ is\footnote{Besides the minimal subtraction scheme used here, other schemes were proposed that make the estimates of $C_{2n}$ and power counting much simpler\cite{ksw,bedaque_bira_review, beane_et_al_review}.}

\begin{eqnarray}
I_0 &=& \left(\frac{\mu}{2}\right)^{3-D} \int \frac{d^{D}q}{(2\pi)^D}\frac{1}{E-\frac{q^2}{M}+i\epsilon}\nonumber\\
&=&-i\frac{M}{4\pi} \sqrt{ME},
\end{eqnarray}where  $D$ is the number of space dimensions and we used the standard relation between the amplitude and the phase shifts in the second line of Equation (\ref{amplitude}). We can now go to the on-shell point $ME=k^2$ and relate the phase shifts to the constants $C_{2n}$ 

\beq\label{master}
k \cot\delta = \frac{4\pi}{M}\frac{1}{\sum_n C_{2n} k^{2n}}.
\eeq

The same combination of low energy constants appearing above also determine the position of the energy eigenstates on a finite volume. They are given by the  poles of the finite volume, real time correlator analogue to Eq.(\ref{correlator})

\begin{eqnarray}\label{corr_singlet}
C(E,\pv) &=&\int dt e^{iEt}\bra{0}T N^T(\pv,t) \Pv N(-\pv,t) \ .\  N^\dagger(-\pv,0) \Pv^{\dagger} N^*(\pv,0) \ket{0}\nn\\
&\sim&\mathop{\sum}_{\vec{q}} \frac{i}{E-E_n+i0} |\bra{E_n}N^\dagger(-\pv,0) \Pv^{\dagger} N^*(\pv,0) \ket{0}|^2 ,
\end{eqnarray}where, in the spin singlet case, $P^A\sim \sigma_2 \tau^2\tau^A$ (the $\sigma$'s ($\tau$'s act on spin(isospin)). The computation of this correlator in the effective theory receives two kinds of contributions: from s-wave interactions and from higher partial wave operators. The first kind are the only ones that survive in the infinite volume limit, if one is careful to either use $\pv=0$ or to average the sink and the source in Equation (\ref{corr_singlet})over all possible directions of $\pv$. 
In a finite volume and with the background field, $\pv$ cannot be zero and cannot be averaged over all directions so higher partial wave interactions, starting with the p-wave, contribute to Equation(\ref{corr_singlet}). These contribution are suppressed by a factor $(Q/m_\pi)^3$ compared to the leading interactions. Furthermore, for $\phi=0,\pi$ this contributions are further suppressed. For $\phi=0$ the cubic symmetry forbids the contamination from p-waves. For $\phi=\pi$ there is an extra ``parity" symmetry along the $z$ axis that, combined with the two dimensional cubic group in the $x-y$ plane also forbids p-wave contributions. As we will see below, values of the background field around $\phi\approx\pi$ are the most interesting ones, so we will disregard the higher partial wave pieces in the following.

\begin{figure}[t]\label{S}
\includegraphics*[
scale=0.8,clip=false]
{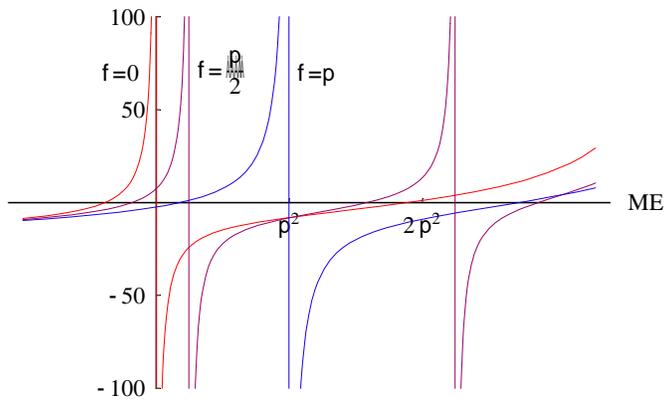}
\caption{\textit{$S(\frac{MEL^2}{4\pi^2},\phi)$  as a function of $ME$ for three values of $\phi=0,\pi/2$ and $\pi$.  } } 
\end{figure}

The correlator in Equation (\ref{corr_singlet}) can be computed in the effective theory and the result is (up to corrections of order $(Q/m_\pi)^3 (\phi-\pi)^2)$)

\beq\label{inverse_amp}
C(E,\kv) \sim \frac{1}{1-\mathop{\sum}_{n} C_{2n} (ME)^{n}\frac{1}{L^3}\mathop{\sum}_{\qv} \frac{1}{E-\frac{q^2}{M}}},
\eeq where the sum is over all the allowed momenta in the box.
For the twisted boundary conditions these allowed momenta are

\begin{eqnarray}
q_{x,y} &=& \frac{2\pi}{L} n_{x,y},\nonumber\\
q_z &=& \frac{2\pi}{L} (n_z+\frac{\phi}{2\pi}),
\end{eqnarray} with $n_x, n_y$ and $n_z$ integers. The poles of Equation (\ref{inverse_amp}) are then determined by

\beq
k L \cot\delta = \frac{1}{\pi} S(\frac{k^2 L^2}{4\pi^2}, \phi),
\eeq with $k^2=ME$ and the function $S(\eta, \phi)$ is defined by

\beq
S(\eta, \phi) \equiv \lim_{N\rightarrow \infty}\sum_{|\vec{n}|<N} \frac{1}{\underline{n}^2-\eta}-4\pi N
\eeq and $\underline{n}\equiv (n_x, n_y, n_z+\phi/2\pi)$. Notice that 
$S(\eta, \pi+\phi)=S(\eta, \pi-\phi)$. In Figure\ref{S} we show $S(\eta,\phi)$ as a function of $\eta$ for a few values of $\phi$.
\footnote{A {\tt C} code computing $S(\eta,\phi)$ in an efficient way can be downloaded from 
{\tt http://www-nsdth.lbl.gov/\~{}bedaque/}.}

For large values of $L$, $kL\cot\delta$ is also large and  Equation (\ref{master}) will be satisfied for values of $k$ where $S$ is close to one of its poles. These poles are located at $k^2 =4\pi^2(n_x^2+n_y^2+(n_z+\phi/2\pi)^2)/L^2= \phi^2/L^2, (4\pi^2+\phi^2)/L^2, \cdots$, corresponding to the eigenstates of free particles in the presence of the background field.
Close to the first pole, for instance, the function $S$ is dominated by the $\vec{n}=0$ term
and we have

\beq
kL\cot\delta=\frac{1}{\pi}\frac{4\pi^2}{\phi^2-k^2L^2}+c_0(\phi)+{\mathcal O}(\left( \frac{\phi^2-k^2L^2}{4\pi^2}\right)^2),
\eeq where $c_0(\phi)$ is

\beq
c_0(\phi) = \lim_{N\rightarrow \infty}\sum_{\stackrel{|\vec{n}|<N}{\vec n \ne 0}} \frac{1}{\vec{n}^2+2 n_z\phi}-4\pi N
\eeq

 The equation above can be solved iteratively determining the  energy level :

\beq
E = \frac{\phi^2}{ML^2} - \frac{4\pi}{kML^3}\tan\delta\left[1+\frac{c_0(\phi)}{kL}\tan\delta  +\cdots\right].
\eeq

For $k\ll m_\pi$, $k\cot\delta$ is well approximated by $k\cot\delta \cong -1/a+r_0 k^2/2 + \cdots$ (effective range expansion), where $a$ is the scattering length and $r_0$ the effective range. For $L\gg \sqrt{ar_0}$, $k\cot\delta\cong -1/a$ and the formula above reduces to

\beq\label{luscher_phi}
E =  \frac{\phi^2}{ML^2} +\frac{4\pi a}{ML^3}\left[1-\frac{c_0(\phi) a}{L} +\cdots \right],
\eeq which is the analogue of the ``Luscher's formula"\cite{hamber_et_al,luscher1,luscher2}.

\begin{figure}
\centering
\subfigure[singlet] 
{
    \label{esinglet}
    \includegraphics[width=0.4\textwidth]{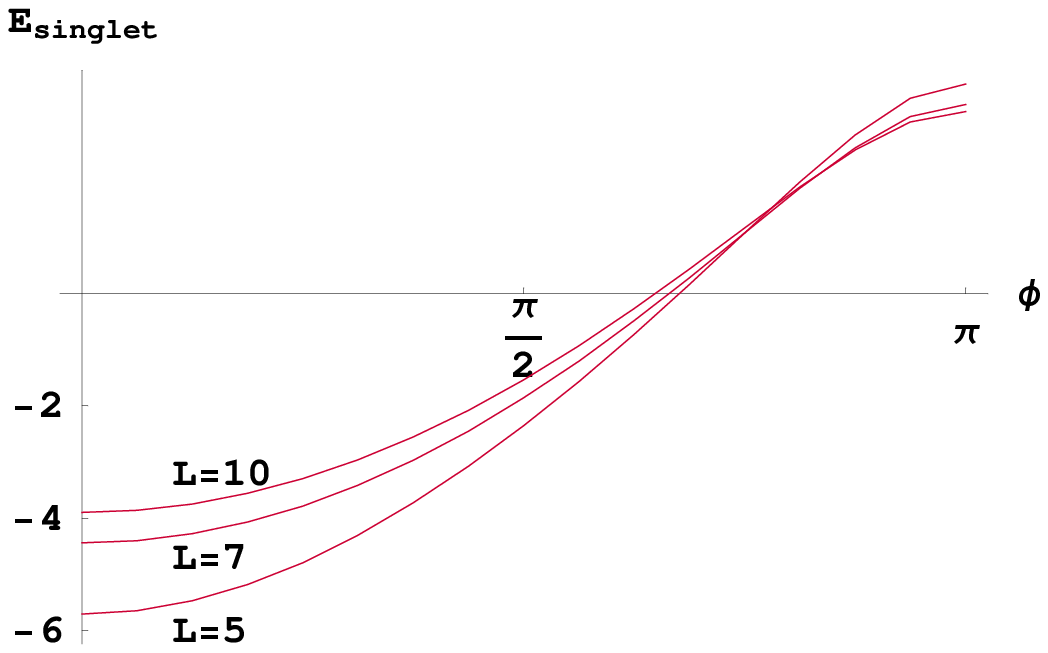}
}
\hspace{1cm}
\subfigure[triplet] 
{
    \label{etriplet}
    \includegraphics[width=0.4\textwidth]{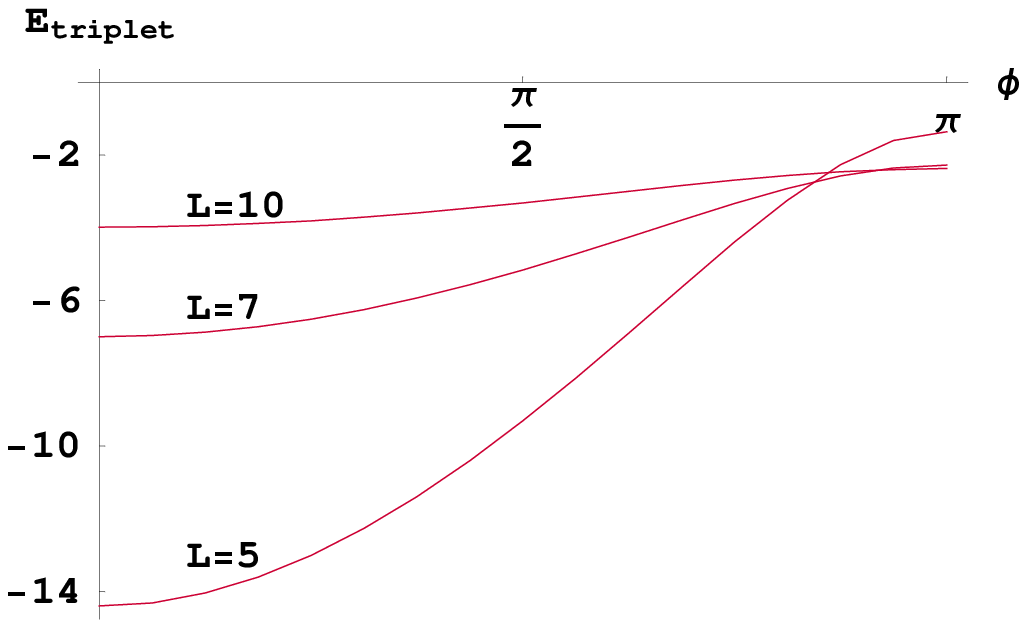}
}
\caption{\textit{Ground state energy (in MeV) of two nucleons in the spin singlet (left) and triplet (right) channel as a function of the background field. The three curves correspond, from bottom to top, $L=5, 7$ and $10$ $fm$. } }
\label{e} 
\end{figure}

For smaller values of $L$ simple expansions as Equation (\ref{luscher_phi}) are not available. Still, we can numerically compute the function $S(\eta,\phi)$ and related energy level in the box with the values of phase shifts. In Figure \ref{esinglet}  we show the estimate of the ground state energy for boxes of different sizes and for different values of the background field. For these estimates we took the values of the phase shifts as given by th effective range formula with parameters $a_s=23.7, r_{0s}=2.73$ for the singlet. 
The validity of our approach is limited to momenta smaller than about half the pion mass, corresponding to an energy scale of about $6$ MeV. Energy states with (the absolute value of the) energy larger than this are an artifact of the effective theory and will not exist in a lattice QCD simulation.

Things are a little more complex in the triplet s-wave channel ($^3S_1$) due to the mixing with the triplet d-wave channel ($^3D_1$). In the infinite volume limit this mixing is generated by the tensor force and is suppressed at low energies by a factor of $(Q/m_\pi)^4$ \cite{chen_pionless}. Two insertions of the tensor force are necessary: one leading from the s-wave to the d-wave and another back to the s-wave. Each one of these transitions is of order $(Q/m_\pi)^2$ and the total effect is $\sim (Q/m_\pi)^4$. In a finite volume, due to the breaking of rotational symmetry by the shape of the torus and the background field, tensor forces can contribute already at leading order. However, it is easy to see by an explicit calculation that the contribution linear in  the tensor force is proportional to $q^iq^j-q^2\delta^{ij}/3$, where $\vec{q}$ is either an internal or external momentum $\vec{q}=\vec{p}$. This contribution vanishes after averaging over spin polarizations. In other words,  the spin averaged correlator

\beq
C(t,\pv) =\frac{1}{3}\sum_{i=1,2,3}\bra{0}T N^T(\pv,t) P^i N(-\pv,t) \   N^\dagger(-\pv,0) P^{i\dagger} N^*(\pv,0) \ket{0},
\eeq ($P^i\sim \sigma^2\sigma^i\tau_2$ is the spin triplet projector) will receive contributions to the tensor force only at second order or higher. Since the contributions of the tensor force at finite or infinite volume are small ($\sim (Q/m_\pi)^4$) we will disregard them here.

In Figure \ref{etriplet}  we show the estimate of the triplet ground state energy for boxes of different sizes and for different values of the background field. We use the values $a_t=5.425$ and $ r_{0t}=1.75$. Again we see that even boxes as small as $L=5$ fm can support a state with energy small enough to be useful in the extraction of phase shifts, if $\phi$ is adjusted to be around $\phi \approx \pi$. In the absence of the background field box sizes of at least $L\approx 8-10$ fm would be required.

In both the spin singlet and triplet cases the lower bound on $L$ saturates the minimum
value set by  finite pion mass effects. These effects are suppressed by the factor $e^{-m_\pi L}$ which
for $L=5$ fm gives $e^{-m_\pi L}\approx0.03$. To use still smaller boxes a similar calculation to the one presented here must be done using an effective theory which includes pions explicitly
and valid for $Q\sim m_\pi \ll m_\rho$. One inconvenient is that calculations in this effective theory will be neccesarily  truncated to a certain order in the low energy expansion,  unlike here where an all orders computation was possible. That means that  the matching with the lattice results will be  more properly described as a computation of the effective theory low energy constants than as a determination of the phase shifts. By the other hand this calculation can be used to extrapolate to realistic values of the quark mass \cite{beane_savage_1, beane_savage_2, meissner_et_al}.

Besides the possibility of using smaller lattice sizes, another advantage of the background field method described here is that the phase shifts at 
arbitrary kinematical points are probed (as long as they are  in the regime $k<m_\pi/2$). 

The method described here may also be applied to other hadronic interactions. For instance, pion-pion phase shifts can be calculated at arbitrary kinematical points by adding a background field coupling to some flavor charge, for instance, electric charge. This spoils the positivity of the determinant so seems to be feasible only in quenched calculations. Similarly, it might be of use  in ameliorating some of the issues involved in the $k\rightarrow \pi\pi$ lattice extraction.

\subsection{Acknowledgments}
The author thanks S. Beane, D. Lin,  A. Parreno, M. Savage and other members of NPLQCD\cite{nplqcd_page} for discussions on the subject.
This work was supported in part by the Director, Office of Energy Research,
  Office of High Energy and Nuclear Physics, by the Office of Basic Energy
  Sciences, Division of Nuclear Sciences, of the U.S. Department of Energy
  under Contract No.~DE-AC03-76SF00098 .


\end{document}